\documentclass[12pt]{article}
\usepackage{graphicx}

\def\CP{$ C \! P$}
\def\ra{\rightarrow}
\def\sbabar{\mbox{{\normalsize \sl B}\hspace{-0.4em} {\small \sl A}\hspace{-0.03em}{\normalsize \sl B}\hspace{-0.4em} {\small \sl A\hspace{-0.02em}R}}}
\let\babar=\sbabar

\def\beq{\begin{equation}}
\def\eeq#1{\label{#1}\end{equation}}
\def\eeqn{\end{equation}}

\def\beqa{\begin{eqnarray}}
\def\eeqa#1{\label{#1}\end{eqnarray}}
\def\eeqan{\end{eqnarray}}

\let\bar=\overbar

\def\Dslash{\not{\hbox{\kern-4pt $D$}}}
\def\dslash{\not{\hbox{\kern-2pt $\del$}}}

\def\msb{{\bar{\ssstyle M \kern -1pt S}}}

\def\Title#1{\begin{center} {\Large {\bf #1} } \end{center}}

\begin{document}

\Title{Exclusion of multifold solutions  of the CKM Unitarity Triangle by a time-dependent Dalitz plot analysis of $\bar B^0  \ra D^{(*)0} h^0$ 
with $D^0 \ra K^0_S \pi^+ \pi^-$ decays  combining  BABAR and Belle data}

\bigskip\bigskip


\begin{raggedright}  

{\it Gerald Eigen (for the \sbabar\ Collaboration)\index{Eigen, G.}\\
Department of Physics and Technology\\
Universisy of Bergen\\
-5007 Bergen,  NORWAY}
\bigskip\bigskip

Talk presented at the APS Division of Particles and Fields Meeting (DPF 2017), July 31-August 4, 2017, Fermilab. C170731.

\abstract{We present results of a new analysis campaign, which combines the final data samples collected by the B factory experiments \sbabar\ and Belle in single physics analyses to achieve a unique sensitivity in time-dependent \CP\ violation measurements. The data samples contain $(471 \pm 3) \times 10^6  ~B \bar B$ pairs recorded by the \sbabar\ detector and $(772 \pm 11) \times 10^6 ~ B \bar B$ pairs recorded by the Belle detector in $e^+e^-$ collisions at the center-of-mass energies corresponding to the mass of the $\Upsilon(4S)$ resonance at the asymmetric-energy $B$ factories PEP-II at SLAC and KEKB at KEK, respectively. We present a measurement of $\sin 2\beta $ and $\cos 2\beta $ by a time-dependent Dalitz plot analysis of $B^0 \ra  D^{(*)} h^0$ with $D \ra K^0_S \pi^+\pi^-$ decays. A first evidence for $\cos 2\beta >0$, the exclusion of trigonometric multifold solutions of the Unitarity Triangle and an observation of \CP\ violation are reported.}

\end{raggedright}

\section{Introduction}

\begin{figure}[htbp!]
\centering
\includegraphics[width=60mm]{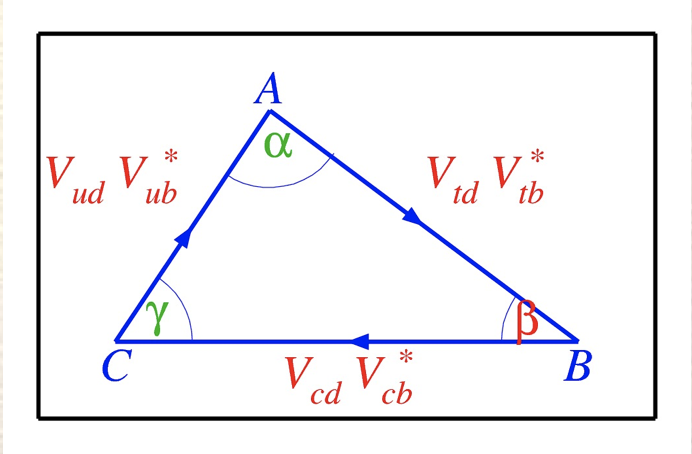}
\includegraphics[width=85mm]{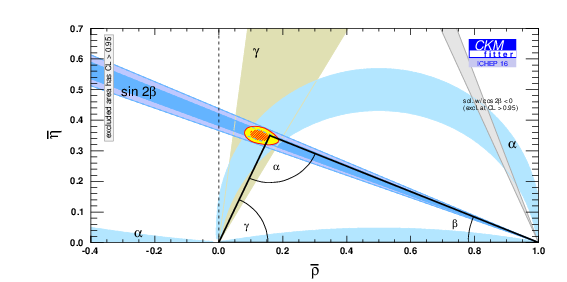}
\caption{Graphical representation of the Unitarity Triangle defining the sides $V_{ub}^* V_{ud}$,  $ V_{cb}^* V_{cd}$, $V_{tb}^* V_{td}$ and angles $\alpha, \beta, \gamma$  (left). Present status of the Unitarity Triangle showing the two solutions of $\beta =21.9^\circ$ and $\beta =68.1^\circ$ (right).  }
\label{fig:UT} 
\end{figure}

In the Standard Model, \CP\ violation originates from the phase of the CKM matrix~\cite{CKM}. The unitarity relation $V_{ub}^* V_{ud} + V_{cb}^* V_{cd} + V_{tb}^* V_{td} =0$ represents the so-called Unitarity Triangle (UT) illustrated in Fig.~\ref{fig:UT} (left). Sides $V_{ub}^* V_{ud}$,  $ V_{cb}^* V_{cd}$, $V_{tb}^* V_{td}$ and angles $\alpha, \beta, \gamma$  are extracted from many measurements in the $B$ and $K$ systems. At the $B$ factories, $\sin 2\beta$ is measured in time-dependent \CP\ asymmetries of $B^0 \ra K^0 c \bar c $ decays in which the $c\bar c$ forms a charmonium resonance~\cite{beta}. 
The determination of $\beta$ from $\sin 2\beta$ measurements leads to a two-fold ambiguity on $ 2\beta$ for $\eta < 90^\circ$, yielding $\beta = 21.9^\circ$ and $\beta = \pi/2 -21.9^\circ=68.1^\circ$ as shown in Fig.~\ref{fig:UT} (right)~\cite{ckmfitter}. To lift this ambiguity, a measurement of $\cos 2 \beta$ is necessary.

\section{Analysis Methodology}

\begin{figure}[htbp!]
\centering
\includegraphics[width=50mm]{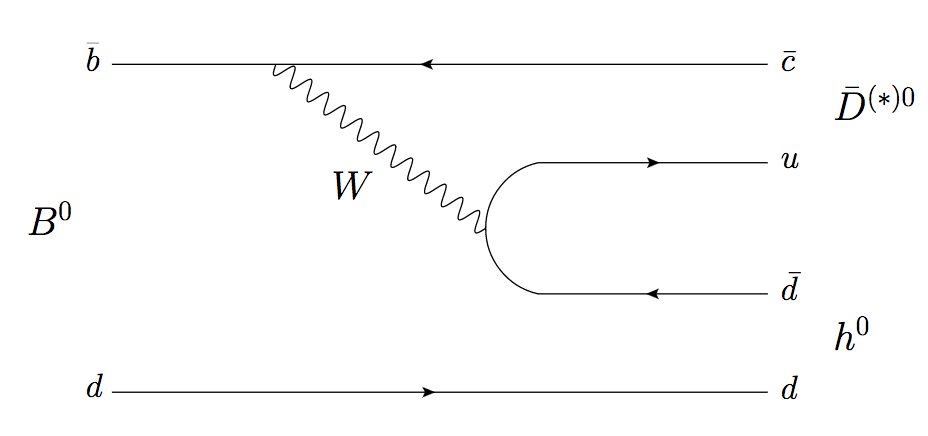}
\includegraphics[width=50mm]{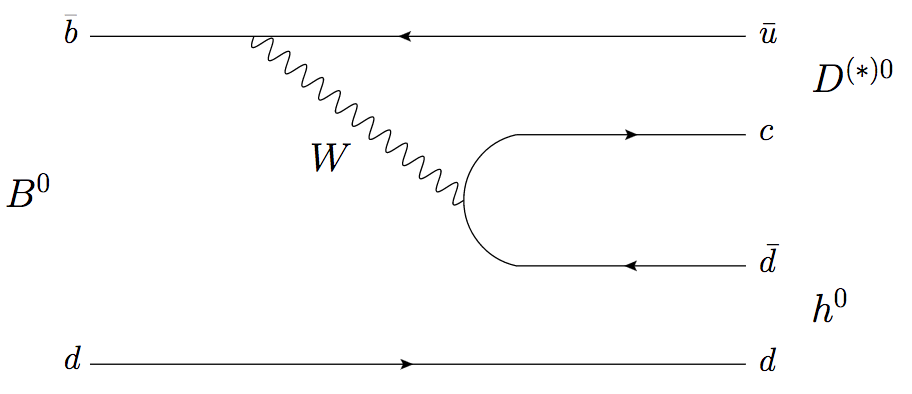}
\caption{Lowest-order tree diagram (left) and doubly-CKM-suppressed tree diagram (right) for $\bar B^0  \ra D^{(*)0} h^0$ with $D^0 \ra K^0_S \pi^+ \pi^-$ decays.}
\label{fig:FD}
\end{figure}

We perform a time-dependent Dalitz plot analysis of $\bar B^0  \ra D^{(*)0} h^0$ with $D^0 \ra K^0_S \pi^+ \pi^-$ decays and where $h^0$ represents a $\pi^0 \ra \gamma \gamma$, $\eta \ra \gamma \gamma, \pi^+ \pi^- \pi^0$ or $\omega \ra \pi^+ \pi^- \pi^0$.  Figure~\ref{fig:FD} shows the lowest-order Feynman diagrams. The $ b \ra c  \bar  u d$ tree dominates whereas 
the contribution from  the doubly-CKM suppressed tree $ b \ra u  \bar  c d$ is small.  The tree amplitudes interferes with the amplitude of   corresponding $\bar B0$ decay that is produced via $B^0 \bar B^0$ mixing. 
Thus, the absolute amplitudes squared for $B^0$ and $\bar B^0$ decay are~\cite{bondar} 

\begin{eqnarray} 
\label{eq:1}
| M_{B^0}(\Delta t) |^2 &=& | {\cal A}_{ \bar D^0} \times \cos (\frac{1}{2} \Delta m \Delta t) - i e^{+2i \beta}\times {\cal A}_{ D^0} \times \sin (\frac{1}{2} \Delta m \Delta t)  |^2 \\  \nonumber
| M_{\bar B^0}(\Delta t) |^2 &=& | {\cal A}_{ D^0} \times \cos (\frac{1}{2} \Delta m \Delta t) - i e^{-2i \beta}\times {\bar \cal A}_{D^0} \times \sin (\frac{1}{2} \Delta m \Delta t) |^2,
\end{eqnarray}
where ${\cal A}_{\bar D^0}$  and  ${\cal A}_{D^0}$ are the  amplitudes of the  $\bar D^0 \ra K^0_S \pi^+ \pi^-$  and $D^0 \ra K^0_S \pi^+ \pi^-$ decays.
The interference of the $\bar D^0$ and the $D^0$ in the Dalitz plot introduce a dependence on $\cos 2 \beta$.
Belle previously measured $\cos 2 \beta = 1.05 \pm 0.33 ^{+0.21}_{-0.15}$~\cite{belle16}. By combining \babar\ and Belle data consisting of a total luminosity of $1.1~\rm ab^{-1}$, the sensitivity to $\cos 2 \beta$ is greatly improved. The first step is to develop a Dalitz plot model for $ D^0 \ra K^0_S \pi^+ \pi^-$ decay followed by the extraction of the $\bar B^0 \ra D^{(*)0 }h^0$ signal for which a time-dependent \CP\ analysis is performed.

\subsection{$D^0 \ra K^0_S \pi^+ \pi^-$ Dalitz Plot Amplitude Model}

\begin{figure}[htbp!]
\centering
\includegraphics[width=80mm]{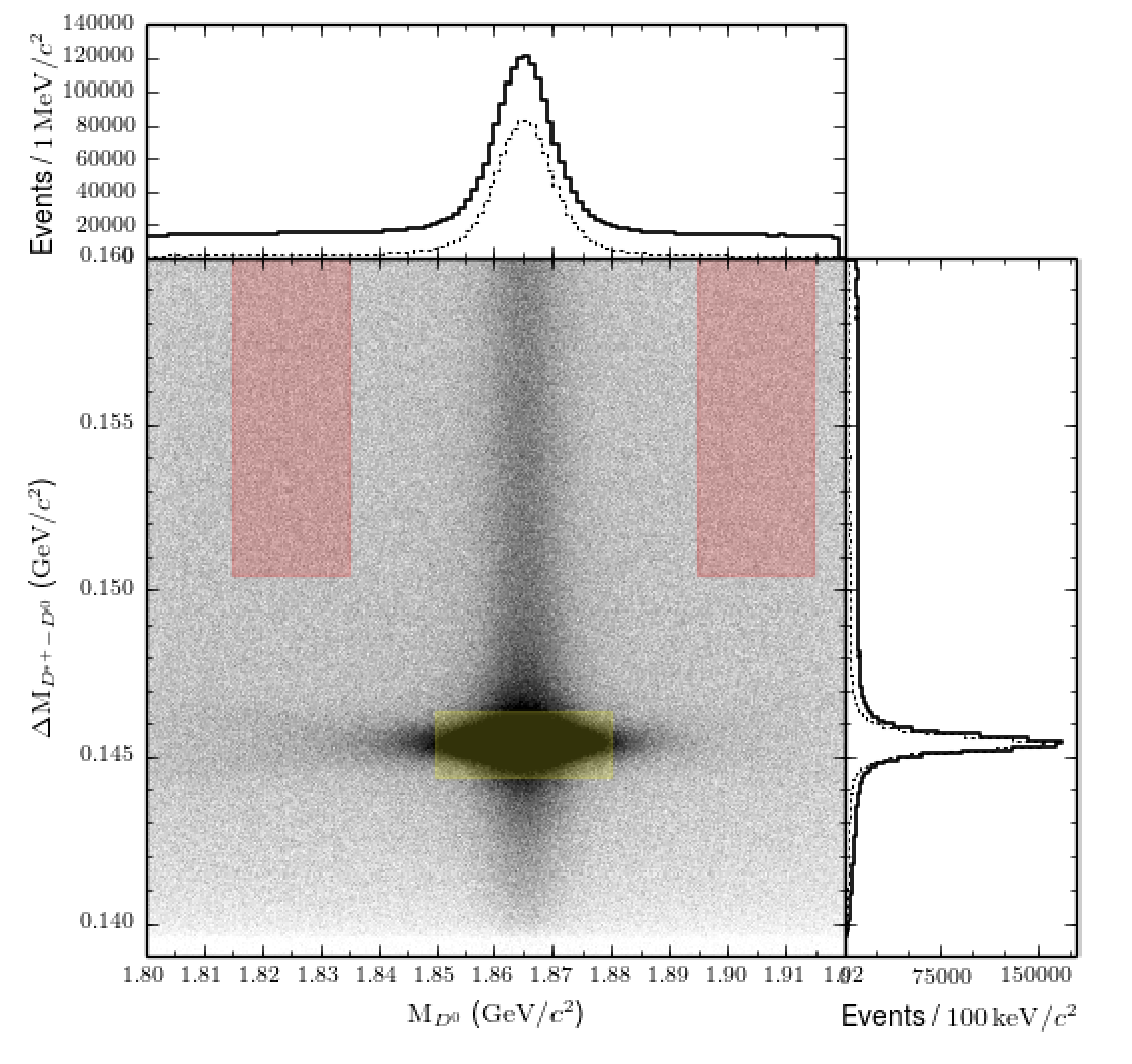}
\caption{Scatter plot of the $D^{*+} -D^0$ mass difference versus the $D^0$ mass and its projections.  The yellow and red bands show the signal and sideband regions, respectively. The dashed lines indicate the signal contributions.  }
\label{fig:D0DstD}
\end{figure}

The $D^0 \ra K^0_S \pi^+ \pi^-$ Dalitz plot model was built using  Belle $e^+e^- \ra c \bar c$ data. The $D^0$ is produced in the decay 
$D^{*+} \ra D^0 \pi^+_s$. The charge of the slow $\pi^+_s$ determines the flavor of the $D^0$. To remove background from $B$ decays, we 
require the candidate momentum for decays in the $\Upsilon(4S)$ ($\Upsilon(5S)$) center-of-mass systems to satisfy $p^*(D^{*+}) >2.5~ (3.1)~\rm GeV$.
We determine $D^{*+}$ decay vertex with a kinematic fit for the $D^0$ daughter with the constraint to originate from the $e^+ e^-$ interaction region. Furthermore, we require that the slow $\pi^+_s$ originates from the $D^{*+}$ decay vertex to improve the  resolution of the $D^{*+} -D^0$ mass difference. We then extract $D^{*+} \ra D^0 \pi^+_s$  yield from a two-dimensional  fit to the $D^0$ mass and $D^{*+} -D^0$ mass difference.  Figure~\ref{fig:D0DstD} shows the scatter plot of the $D^0$ mass versus the $D^{*+}-D^0$ mass difference. We obtain a yield of
of 1.22 million signal events in the region of $144.4 < M_{D^*}-M_D < 146.4$ MeV and  $1850 < M_{D^0} < 1880$ MeV with a purity
of $94\%$.

\begin{figure}[htbp!]
\centering
\includegraphics[width=100mm]{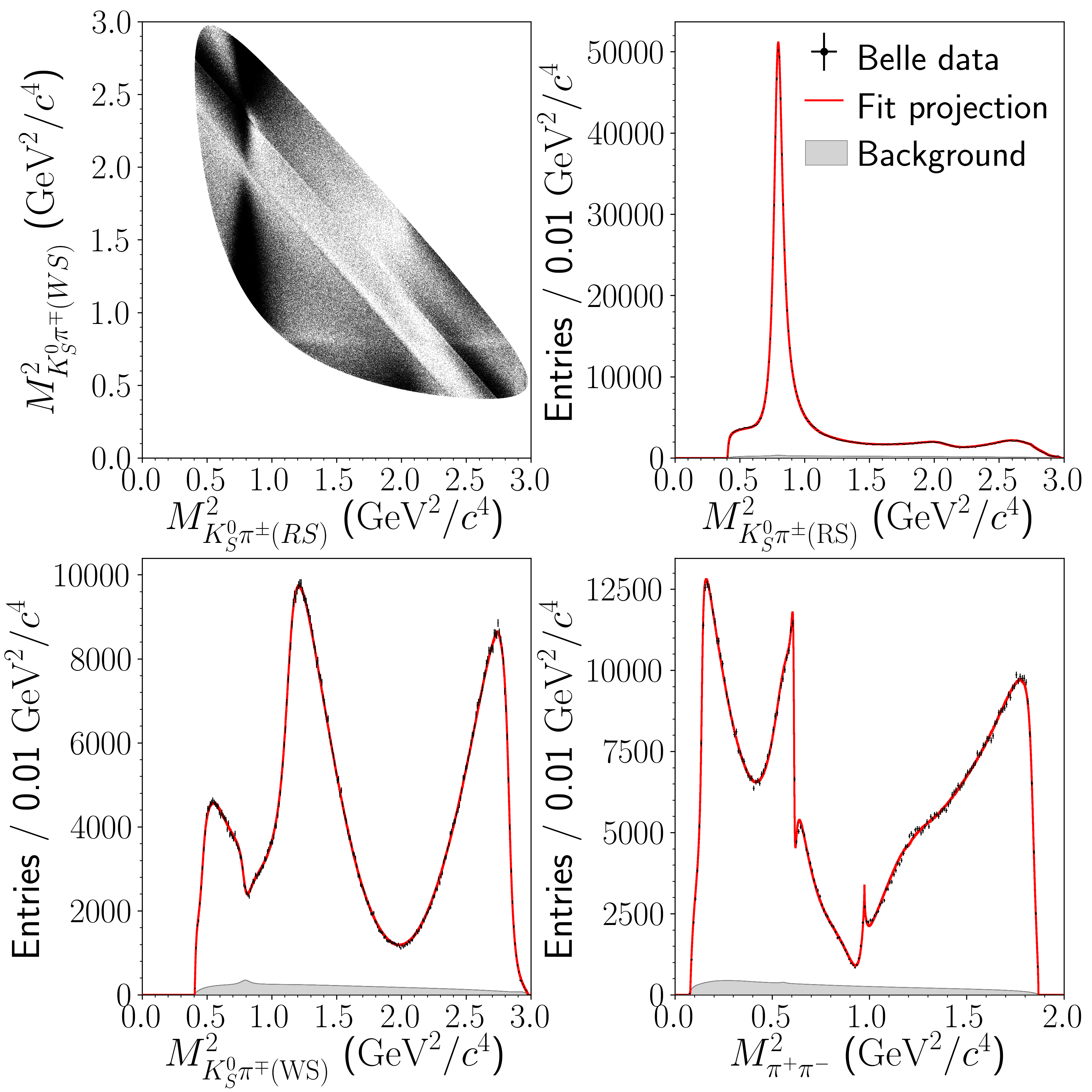}
\caption{The  $M^2_{K^0_S\pi^\pm}$ versus  $M^2_{K^0_S \pi^\mp}$ Dalitz plot and the  $M^2_{K^0_S \pi^\pm}$,   $M^2_{K^0_S \pi^\mp}$ and  $M^2_{\pi^+ \pi^-}$ projections for Belle data with fits results overlaid. The residual $6\%$ $c \bar c$ background is represented by the grey-shaded distribution.}
\label{fig:D0DP}
\end{figure}

We fit  the $D^0 \ra K^0_S \pi^+ \pi^-$ Dalitz plot with isobars accounting for 13 intermediate two-body resonances and  $K \pi$   and $ \pi \pi$ S-waves 
where the S-waves are  modelled by the LASS~\cite{lass}  and K-matrix~\cite{chung} parameterizations, respectively. 

\begin{equation}
{\cal A}_{D^0} (m^2_+, m^2_-) =\sum_{r \ne (K\pi,\pi \pi)_{\rm L=0}} a_r e^{i \phi_r} {\cal A}_r (m^2_+, m^2_-)+ {\cal A}_{(k \pi)_{\rm L=0}} (s) + F_1(s)
\end{equation}
\noindent
The intermediate resonances are parametrized in terms of manitude $a_r$, phase $\phi_r$ and amplitude ${\cal A}_r(m^2_+, m^2_-)$. The latter is factorized in terms  of the Zemach tensor $(Z_L)$~\cite{zemach} describing the angular distribution of final state, Blatt-Weisskopf barrier penetration factors for the $D $ meson $(F_D)$ and the  resonance $(F_r)$ as well as a propagator $(T_r)$ describing the decay dynamics of the resonances parameterized by Breit-Wigner functions. The isobar model includes Cabibbo-allowed two-body resonances  ($ K^*(890)^-$, $K^*_0(1430)^-$, $K^*_2(1430)^-$, $ K^*(1680)^-$ and $K^*(1410)^-$), doubly-CKM-suppressed states ($ K^*(890)^+,$ $K^*_0(1430)^+$, $K^*_2(1430)^+$, and $K^*(1410)^+$) as well as \CP~ eigenstates ($\rho(770)$, $\omega(782)$, $f_2(1270)$ and $\rho(1450))$. We perform a Dalitz plot fit for flavor-tagged events in the signal region with a correction for efficiency variations in Dalitz plot phase space. We take background from the  $D^0$ mass and $D^{*+} -D^0$ mass difference sidebands. The free parameters in the fit are $a_r$ and $\phi_r$ of each resonance relative to those of the $\rho$ ($a_\rho=1$ and $\phi_r=0$), the LASS  and $K$-matrix parameters plus masses and widths of the $K^*(892)$ and $K^*(1430)$. 
Figure~\ref{fig:D0DP} shows the Dalitz plot and its three projections  for the Belle $c \bar c$ sample with  fit results overlaid. The $K^*(892)$ is the dominant resonance. It is clearly visible  in the Cabibbo-allowed Dalitz plot projection and appears as reflections at $1.2 ~\rm GeV^2$ and $2.5 ~\rm GeV^2$ in the doubly-CKM suppressed projection. 

\subsection{Extraction of the $\bar B^0 \ra D^{(*)0} h^0$ signal yield}

In total, we reconstruct  five $\bar B^0$ decay modes $(D^0 \pi^0, D^0 \eta, D^0 \omega, D^{*0} \pi^0$ and $D^{*0}\eta$) in which the $D^0$, $D^{*0}$ and neutral hadron $h^0$ are reconstructed in the decays $D^0 \ra K^0_S \pi^+ \pi^-$, 
 $D^{* 0} \ra   D^0\pi^0$, and  $(\eta^0 \ra \gamma \gamma$, $\eta \ra \gamma \gamma$ or  $\eta \ra \pi^+ \pi^- \pi^0$ and $\omega \ra \pi^+ \pi^- \pi^0$).
We introduce a transformed beam-constrained mass, $M_{\rm bc}^\prime $, to remove the correlation between the energy difference $\Delta E=E^*_B - E^*_{\rm beam}$\footnote{where $E_b^*$ is the B-meson energy and $E_{\rm beam}^*$ is the beam energy where an asterisk denotes that these observables are measured in the $e^+ e^-$ center-of-mass frame} and $M_{\rm bc} $ 
\begin{equation}
M_{\rm bc}^\prime= \sqrt{E^{*2}_{\rm beam}-\big( \vec p^*_{D^{(*)0}} +\frac{\vec p^*_{h^0}}{|\vec p^*_{h^0} |}  \sqrt{(E^*_{\rm beam}-E^*_{D^{(*)0} )2-M^2_{h^0}      }\big)}}.
\end{equation}
To separate signal from $e^+ e^- \ra q \bar q $  continuum background (with $q =u,~ d,~ s,~c$ quarks), we define a transferred neural network output variable
\begin{equation}
NN^\prime_{\rm out} = \log \frac{NN_{\rm out} - NN^{min}_{\rm out}}{NN^{max}_{\rm out} - NN_{\rm out}}
\end{equation}

\begin{figure}[htbp!]
\centering
\includegraphics[width=95mm]{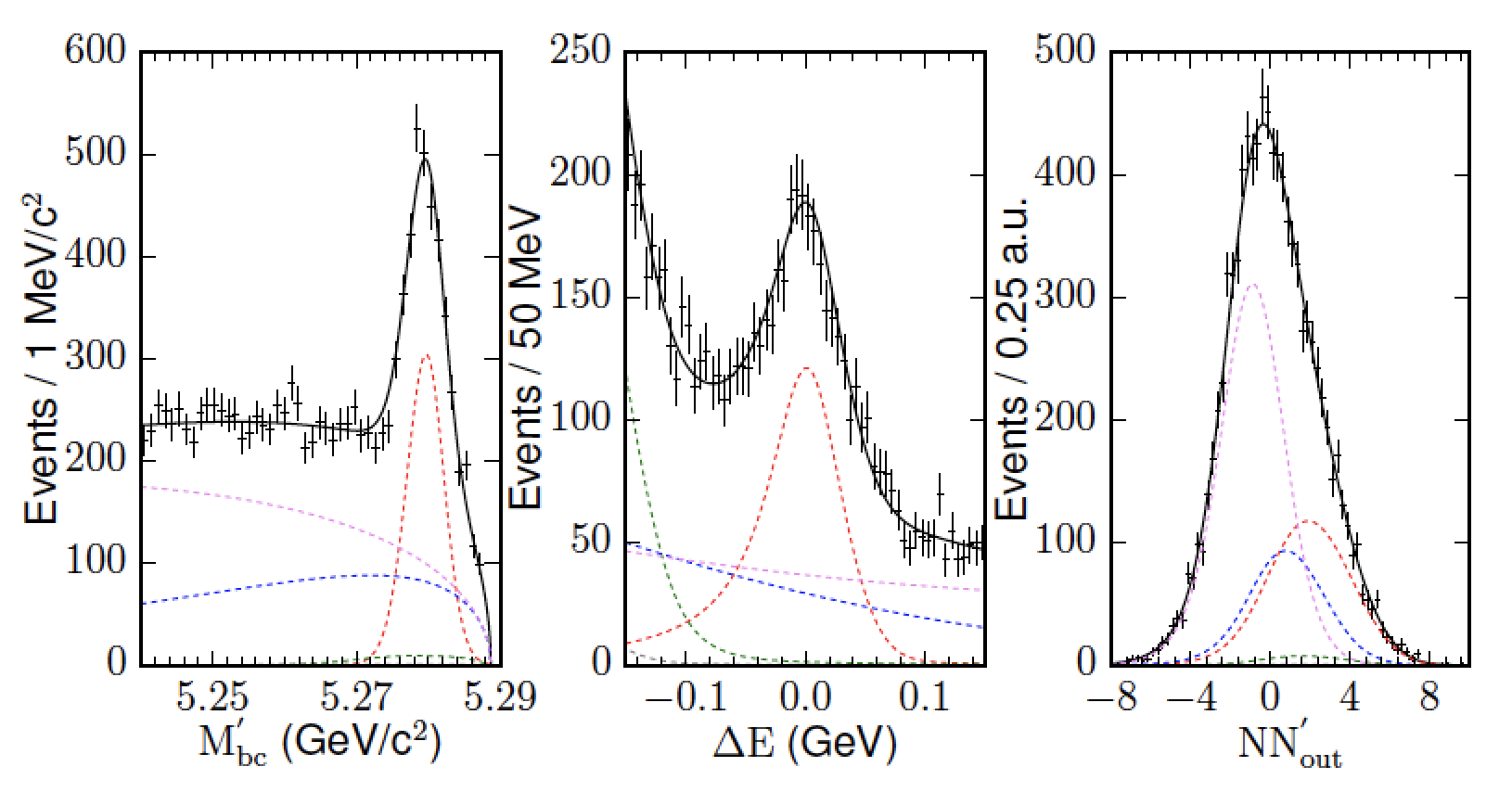}
\vskip -0.2cm
\caption{Distributions of $M^\prime_{\rm bc}$ (left), $\Delta E$ (middle) and $NN^\prime_{\rm out}$ (right) showing data (points), total fit (black line), signal (red dashes), combinatorial background (blue dashes), continuum background (magenta dashes) and cross-feed (green dashes). }
\label{fig:fitprojection}
\end{figure}

The neural network variable $NN_{\rm out}$ combines event shape information from 16 modified Fox-Wolfram moments. We perform a coherent analysis strategy, applying essentially the same selection criteria on \babar\ and  Belle data.
We extract signal yields from three-dimensional fits to $M^\prime_{\rm bc}$, $\Delta E$ and $NN^\prime_{\rm \rm out}$. For each mode in each experiment we perform separate fits using the experiment-specific resolutions. Figure~\ref{fig:fitprojection} shows the distributions for  $M^\prime_{\rm bc}$, $\Delta E$ and $NN^\prime_{\rm out}$ with  fit projections overlaid after combining all five modes for both experiments. The three-dimensional fits describe the data rather well. We obtain $1129 \pm 48$ \babar\  and $1567 \pm 56$ Belle signal events. Table~\ref{tab:yields} lists the yields for the individual modes in the two experiments.
 
\begin{table}[htbp!]
\begin{center}
\begin{tabular}{l|ccc}  
Decay mode &  \babar &   Belle \\ \hline
$\bar B^0 \ra D^0 \pi^0$ &   $469\pm 31$       &   $768\pm 37$  \\
$\bar B^0 \ra D^0 \eta$ &   $220\pm 22$       &   $238\pm 23$  \\
$\bar B^0 \ra D^0 \omega$ &   $219\pm 21$       &   $285\pm 26$  \\
$\bar B^0 \ra D^* \pi^0$ &   $147\pm 18$       &   $182\pm 19$  \\
$\bar B^0 \ra D^* \eta$ &   $74\pm 11$       &   $94\pm 13$  \\ \hline
\end{tabular}
\vskip -0.2cm
\caption{Individual signal yields per decay mode and experiment.}
\label{tab:yields}
\end{center}
\end{table}

\section{Time-dependent \CP\ analysis}                                                                                                         

We fit the proper time interval distributions maximizing the log-likelihood function

\begin{equation}
\ln {\cal P}= \sum_i \ln {\cal P}_i^{BABAR} + \sum_j \ln {\cal P}_j^{Belle}.
\end{equation}
\noindent
The physics probability density functions (pdf)  ${\cal P}_i^{BABAR}$ and ${\cal P}_j^{Belle}$  are convolved with experiment-specific resolution functions

\begin{equation}
{\cal P}_{\rm exp} =\sum_k f_k \int [{\cal P}_k(\Delta t) R_k(\Delta t- \Delta t^\prime)] d\Delta t^\prime .
\end{equation}
\noindent
We apply the \babar- and  Belle-specific flavor-tagging algorithms and a common signal model for both experiments. The signal pdf is obtained from
eqn~\ref{eq:1}.
\begin{eqnarray}
{\cal P}_{\rm sig}& \propto & [|{\cal A}_{\bar D^0} |^2 +  |{\cal A}_{ D^0} |^2 ] \mp [|{\cal A}_{\bar D^0} |^2  - |{\cal A}_{ D^0} |^2 ] \cos (\Delta m \Delta t) \\ \nonumber 
&\pm &2 \eta_{h^0} (-1)^L  [ Im ({\cal A}_{D^0}  {\cal A}^*_{D^0})  \cos 2 \beta -Re ({\cal A}_{D^0}  {\cal A}^*_{D^0})  \sin 2 \beta  ] \sin (\Delta m \Delta t)
\end{eqnarray}

The $B^0$ and $B^+$ lifetimes and the $B^0 \bar B^0$ mixing parameter $\Delta m_d$ are fixed to the world averages~\cite{pdg}, while the
Dalitz plot amplitude model parameters are fixed to the fit results of the $D^0 \ra K^0_S \pi^+ \pi^Ð$ Dalitz plot analysis.  The only free parameters in the fit are $\sin 2 \beta$ and $\cos 2\beta$. 


\section{Results}     

Using $1.1~\rm  ab^{-1}$ of \babar\ and Belle data we measure:
\begin{eqnarray}
\sin 2 \beta & = 0.80 \pm 0.14_{\rm stat} \pm 0.06_{\rm sys} \pm +0.03_{\rm model} \\ 
\cos 2 \beta & = 0.91 \pm 0.22_{\rm stat} \pm 0.09_{\rm sys} \pm +0.07_{\rm model} \\
\beta & = ( 22.5 \pm 4.4_{\rm stat} \pm 1.2_{\rm sys} \pm 0.6_{\rm model})^\circ
\end{eqnarray}

\begin{figure}[htbp!]
\centering
\includegraphics[width=74mm]{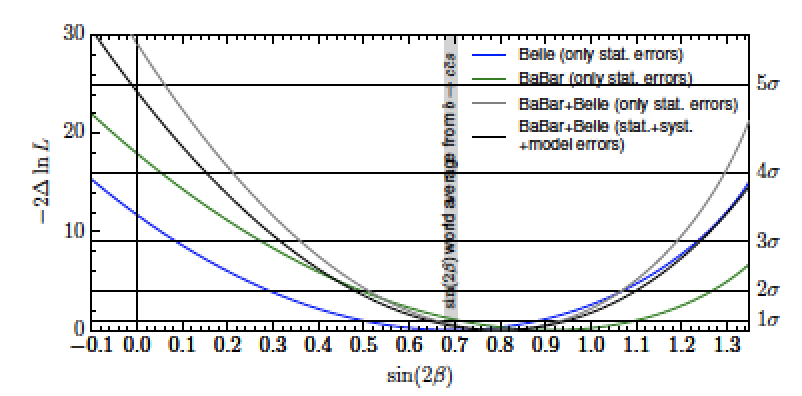}
\includegraphics[width=74mm]{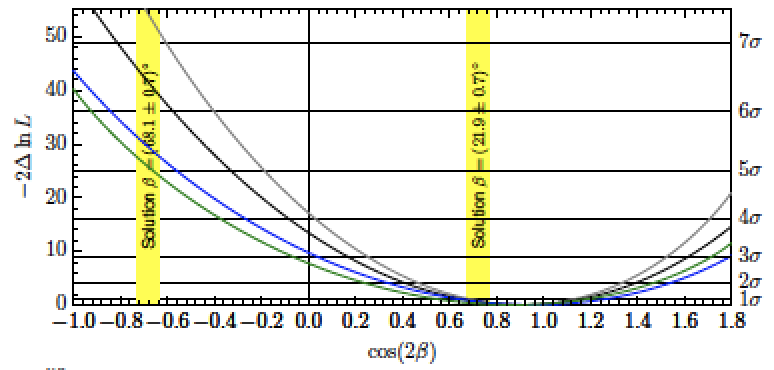}
\includegraphics[width=74mm]{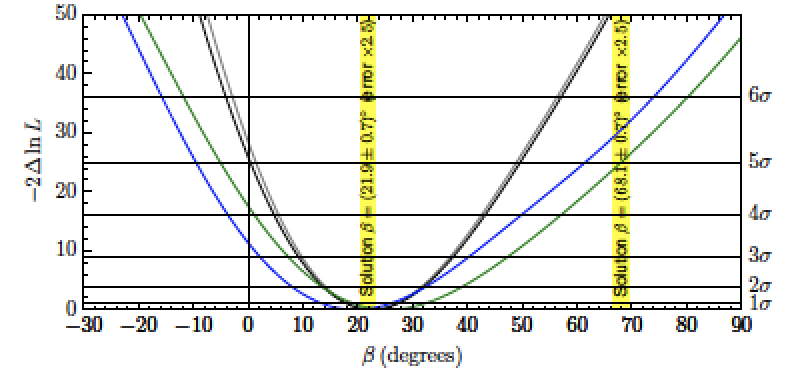}
\caption{ The $2 \Delta \ln L$ distributions as functions of $\sin 2 \beta$ (top left), $\cos 2 \beta$ (top right) and $\beta$ (bottom) for \babar\ data (green line), Belle data (blue line) and both experiments combined (grey line) using statistical errors only. The result for both experiments including systematic uncertainties are shown by the black solid curves. }
\label{fig:likelihood}
\end{figure}

We observe \CP\ violation in the decay $\bar B \ra D^{(*)} h^0$ in the time-dependent decay distributions as expected in the Standard Model. 
The value of $\sin 2 \beta $ is in good agreement with the world average of $0.69 \pm 0.02$~\cite{HFLAV}. Our results yield the first evidence for $\cos 2 \beta >0$ at  3.7 standard deviations. The second solution of $\beta =(68.1 \pm 0.7)^\circ$ is excluded at 7.3 standard deviations thus removing the ambiguity in extracting $\beta$ from $\sin 2 \beta$. In addition, $\beta = 0$ is excluded at 5.1 standard deviations. 
Figure~\ref{fig:likelihood} shows the $2 \Delta \ln L$ distributions as functions of $\sin 2 \beta$, $\cos 2 \beta$ and $\beta$.

\section{Systematic uncertainties}     

Table ~\ref{tab:systematics} shows the experimental systematic uncertainties on the \CP\ violation parameters. The largest  systematic errors on $\cos 2\beta$ result from the uncertainties in the  $\Delta t$ resolution functions (0.058), followed by those in the vertex reconstruction, possible fit bias and signal purity. Systematic uncertainties from the background $\Delta t$  pdfs are smaller while those from  flavor tagging, physics parameters and Dalitz plot reconstruction efficiency corrections are  negligible. 

\begin{table}[htbp!]
\begin{center}
\begin{tabular}{l|ccccc}  
Source &  $\sin 2 \beta [\times 10^2]$ &   $\cos 2 \beta [\times 10^2]$ & $\beta [^\circ]$ \\ \hline
$ \Delta t$ resolution functions &   2.84      &  5.75  & 0.41  \\
Vertex reconstruction &   3.16     &  4.79  & 0.53  \\
Possible fit bias  &   3.67      & 3.90  & 0.79  \\
Signal purity &   2.13      & 3.39  & 0.53  \\
Background $\Delta t$ pdfs &   1.24     & 1.76  & 0.16  \\
Flavor tagging&   0.34      &  0.39  & 0.07  \\
Dalitz plot reconstruction efficiency correction &   0.01      &  0.17  & 0.02  \\
Physics parameters &   0.07      &  0.14  & 0.02  \\  \hline 
Total & 6.14 & 9.27 & 1.18 \\  \hline \hline
\end{tabular}
\caption{Summary of experimental systematic uncertainties on the \CP\ violation parameters.}
\label{tab:systematics}
\end{center}
\end{table}

\section{Conclusions}     

\begin{figure}[htbp!]
\centering
\includegraphics[width=80mm]{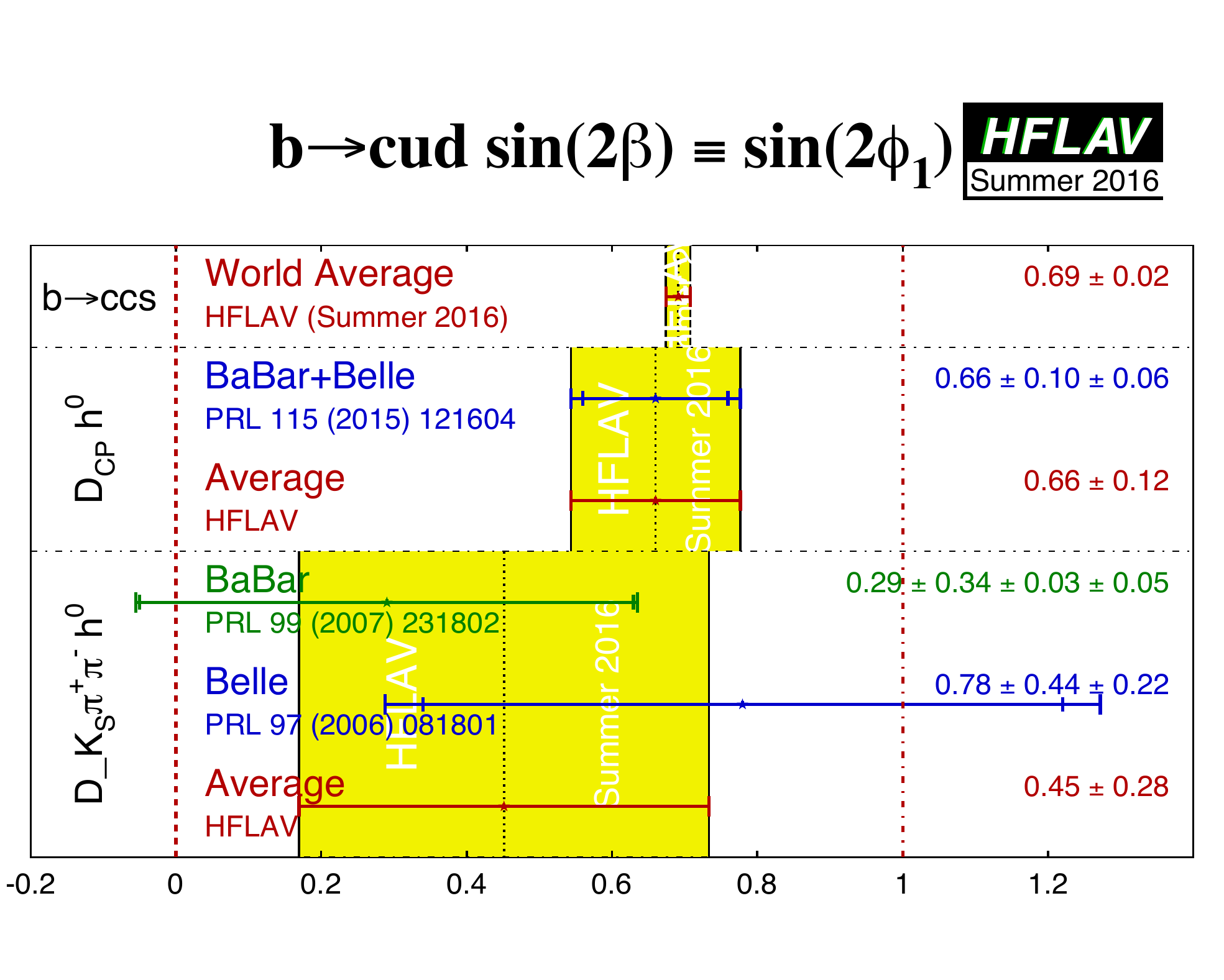}
\caption{Compilation of $\sin 2 \beta^{eff}$ measurement in time dependent $b \ra c \bar u d$  analyses before the resent mmeasurement~\cite{BB15, babar07, belle06}.  }
\label{fig:beta}
\end{figure}

We have performed a time-dependent \CP\ analysis of the decay mode $\bar B^0 \ra D^{(*)0} h^0$ where the $D^0$ is reconstructed in the $K^0_S \pi^+ \pi^-$ final state using $1.1 ~\rm ab^{-1}$ of \babar\ and Belle data. We measure $\cos 2 \beta = 0.91 \pm 0.22_{\rm stat} \pm 0.09_{\rm sys} \pm 0.07_{\rm model}$ 
 from which  we  exclude the solution $\beta =(68.1 \pm 0.7)^\circ$ at 7.3 standard deviations. Our measurement lifts the ambiguity in the determination of  the angle $\beta$ in the  Unitarity Triangle. Our result for $\beta$ agrees well with the world average which is shown in Fig.~\ref{fig:beta} in comparison to a previous measurement of $\beta$ in the decay mode $\bar B^0 \ra D_{CP} h^0$ using combined \babar\ and Belle data. 

\section{Acknowledgment} 

I would like to thank the \sbabar\ collaboration  for  the opportunity to present these results. In particular I would like to thank M. R\"ohrken,  J. McKenna and F.C. Porter for useful discussion.

\end{document}